\documentclass[showpacs,preprintnumbers,amsmath,amssymb,twocolumn,prb]{revtex4}
\usepackage{amssymb,amsmath}
\usepackage{graphics}
\usepackage{epstopdf}
\usepackage{epsfig}
\usepackage{color}
\usepackage{amsfonts}
\usepackage{bm}
\usepackage{amscd}

\begin{document}

\title{General analytical solution to exact fermion master equation}

\author{Matisse Wei-Yuan Tu, Jian-Heng Liu, and Wei-Min Zhang}
\affiliation{Department of Physics and Centre for Quantum Information Science, National Cheng Kung University,
Tainan 70101, Taiwan}


\begin{abstract}
The exact fermion master equation previously obtained in [Phys. Rev. B \textbf{78}%
, 235311 (2008); New J. Phys. \textbf{12}, 083013 (2010)] describes the dynamics of quantum states of a principal system of fermionic particles under the influences of external fermion reservoirs (e.g. nanoelectronic systems). Here, we present the general solution to this exact fermion master equation. The solution is analytically expressed in the most intuitive particle number representation. It is applicable to an arbitrary number of orbitals in the principal system prepared at arbitrary initial states. We demonstrate the usefulness of such general solution with the transient dynamics of nanostructured artificial molecules. We show how various initial states can lead to distinct transient dynamics, manifesting a multitude of underlying transition pathways.
\end{abstract}


\pacs{73.23.-b, 73.63.-b}

\keywords{Quantum decoherence, open quantum systems, quantum dots,
Aharonov-Bohm effect} \maketitle

\section{Introduction}
Nanoelectronic systems are versatile platforms to explore quantum device applications. Conventionally, the Schwinger-Keldysh nonequilibrium formalism is applied to study their quantum transport properties,\cite{Haug08book} where the quantum states of the devices are not necessarily tackled. On the other hand, for the interests of quantum information processing, directly addressing the quantum states of device systems has been fundamentally important.\cite{Nielsen00book} Many experimental endeavors have gone to tailor and manipulate quantum states of the targeted devices.\cite{Ladd1045,Stajic131163} Quantum devices are open systems where interactions with environments can significantly affect the states of the device.\cite{Weiss99book,Zurek03715,Zhang12170402} Equations of motion that govern the time evolutions of the states of open systems, known as master equations, then become common theoretical tools. To have complete characterization of the device operations, solutions to master equations, namely, explicit expressions of the quantum states in terms of the reduced density operators, are strongly sought for.\cite{Fano5774,Schoeller9418436,Gurvitz9615932,Jin08234703,Tu08235311,Jin10083013}

Preparing as well as reading quantum states have been experimentally demonstrated with atoms, ions and molecules,\cite{Leibfried964281,Kurtsiefer97150,Dunn95884,Waxer97R2491,Blatt081008,Volz06692,Mueller07200405,Bloch12267} superconducting circuits,\cite{Nakamura99786,Chiorescu031869,Hofheinz09546,Wang09200404,Filipp09200402,Shalibo13100404,You11590,Xiang13623} photonic modes,\cite{Smithey931244,Schiller962933,Breitenbach97471,Lvovsky11050402,Ourjoumtsev06213601,Deleglise08510,Lvovsky09299} and electrons in quantum dots (QDs),\cite{Hayashi03226804,Elzerman03161308R,Gorman05090502,Petta052180,Fujisawa06759,Hanson071217,Petersson102789,Fricke13126803,Shi143020} just to name a few. Reconstructing quantum states from experimental data of measurements of observables, known as quantum state tomography, have also been intensively investigated on the physical and algorithmic aspects.\cite{Paris04book,Liu04874,Lobino08563,Cramer10149,Baumgratz13020401,Ferrie14190404} On the intersections of quantum information processing, nonequilibrium phenomena and nanoscale electronics lies a class of nanoelectronic systems. They consist of a central system containing a number of discrete states, exchanging electrons with electron reservoirs, as electrodes or other continua in host materials. Sophisticated nano-fabrication technology makes these systems plainly available and highly tunable. Alternatively, using cold atoms with optical controls, fermionic atoms with discretized montional stateas in a trap, connecting to macroscopic fermion reservoirs, have been realized experimentally as analogs to the aforementioned nanoelectronic systems.\cite{Brantut121069,Brantut13713,Krinner1564} Their atomic transport properties have also raised much theoretical interests.\cite{Bruderer12013623,Nietner14013605,Gallego-Marcos14033614} Therefore, exploiting the quantum states of these systems for both purposes of device-oriented and fundamental studies is very attractive.

In addition to the Schwinger-Keldysh formalism, master equation approaches have also been widely applied in nanoelectronic device systems as well.\cite{Schoeller9418436,Gurvitz9615932,Jin08234703,Tu08235311,Li05205304,Harbola06235309}
Perturbations on tunnelings to leads are often employed when Coulomb interactions within the central region are considered.\cite{Schoeller9418436,Li05205304,Harbola06235309} For effectively non-interacting fermions, exact master equations have been obtained\cite{Tu08235311,Jin10083013} via the Feynman-Vernon's influence functional approach\cite{Feynman63118} and strong Coulomb repulsions can be taken into account by excluding double occupancies via modifying the rate equations.\cite{Tu08235311} It is known that the coefficients in the master equations reveals information of decoherence properties.\cite{Carmichael93book,Breuer02book,Caldeira83587,Tu08235311,Lei121408,Zhang12170402} However, to have direct access to the quantum states, solutions to the master equations are demanded. Solving the master equations in real time for the full non-Markov dynamics is generally challenging. Here we present the general analytical solution to the exact master equations for effectively non-interacting fermions. We provide direct and complete descriptions of the quantum states in terms of the reduced density operator in the Fock basis.


As a demonstration, we apply this general solution to the study of transient dynamics of nanoscale artificial molecules. This could be a double-well trap for fermionic atoms or a coupled double-quantum-dot (DQD). For such two-orbital systems, we have previously obtained the analytical solution only under the condition that the initial state is an empty state.\cite{Tu11115318} Here we have arrived at analytical solutions that enable us to explore different transition pathways arising from different initial states.


The organization of this article is the following. In Sec.~\ref{sec_model}, we first briefly review the exact master equation for a class of nanoelectronic systems and their alternative realizations using cold fermion atoms. Our main result, the analytical expression of the reduced density matrix that specifies the quantum state of the principal system, is prescribed in Sec.~\ref{sec_sol}. In Sec.~\ref{sec_DQDmodel}, we illustrate the utilization of such solution with the transient dynamics of artificial molecules. A summary is made in Sec.~\ref{sec_conclu}.




\section{Exact fermion master equations and their general analytical solution}

\subsection{Exact fermion master equations}
\label{sec_model}
We consider the following general Hamiltonian for fermions,
\begin{subequations}
\label{gel_H}
\begin{equation}
\mathcal{H}(t)=\mathcal{H}_{\text{S}}(t)+\mathcal{H}_{\text{E}}(t)+\mathcal{H}_{\text{T}}(t),
\end{equation}
where
\begin{align}
\label{gel_Hs}
\mathcal{H}_{\text{S}}(t)=\sum_{i,j\in\text{S}}\varepsilon_{ij}(t)a_{i}^{\dagger}a^{}_{j},
\end{align} is the Hamiltonian for the principal system, and
\begin{align}
\label{gel_HE}
\mathcal{H}_{\text{E}}(t)=\sum_{\alpha}\sum_{k\in\alpha}\varepsilon_{\alpha k}(t)c_{\alpha k}^{\dagger}c^{}_{\alpha k},
\end{align} is the sum of Hamiltonians of fermion reservoirs, each labeled by $\alpha$.
The exchange of particles between the principal system and the reservoirs is described by
\begin{align}
\label{gel_HT}
\mathcal{H}_{\text{T}}(t)=\sum_{i,\alpha k}\left\{ V_{\alpha ki}(t)c_{\alpha k}^{\dagger}a^{}_{i}+V_{i\alpha k}(t)a^{\dagger}_{i}c_{\alpha k}^{}\right\}
\end{align}
\end{subequations} Here the subscript $i\in\text{S}=\{1,2,\cdots,D\}$ enumerates the orbitals (discrete states) within the principal system and $k\in\alpha$ stands for the continuum states within that reservoir $\alpha$. The fermion field operator $a_{i}(a^{\dagger}_{i})$ or $c^{}_{\alpha k}(c^{\dagger}_{\alpha k})$ then annihilates (creates) a particle on orbital $i\in\text{S}$  or state $k\in\alpha$. The hopping amplitude between an orbital $i$ in the principal system and a state $k$ in reservoir $\alpha$ is given by $V_{\alpha ki}(t)=V^{*}_{i\alpha k}(t)$. The indices $i$ and $k$ include both external (motional) as well as intrinsic (spin or pseudo-spin) degrees of freedom of the fermion in question. The time-dependence of the Hamiltonian parameters can arise from externally applied fields (via gate-voltages in the context of nanoelectronics or laser fields in the context of their cold-atom analogs).

The quantum state of the principal system as an open system is completely given by the reduced density operator, defined as
\begin{equation}
\label{def_RDOPT}
\hat{\rho}_{}\left(t\right)=\text{tr}_{\text{E}}\hat{\rho}_{\text{tot}}\left(t\right),
\end{equation}
where $\hat{\rho}_{\text{tot}}\left(t\right)$
is the density operator of the whole system at time $t$ and $\text{tr}_{\text{E}}$
means tracing over all reservoir degrees of freedom.


We assume as usual\cite{Leg871,Carmichael93book,Weiss99book,Breuer02book} that the total system at initial time $t=t_0$ is in a state $\hat{\rho}_{\text{tot}}(t_0)$
which is a disentangled product of thermal equilibria of the separate
reservoirs, each with its own initial chemical potential $\mu_{\alpha}$ and temperature $T_{\alpha}$, namely,
\begin{subequations}
\label{init_eqthm}
\begin{align}
\hat{\rho}_{\text{tot}}^{}\left(
t_0\right)=\hat{\rho}^{}_{}(t_0)\otimes\prod^{\otimes}_{\alpha}\hat{\rho}^{}_{\alpha}(t_0),
\end{align} where
\begin{align}
\hat{\rho}^{}_{\alpha}(t_0)=\frac{\exp\left[-\left(\mathcal{H}^{}_{\alpha}(t_0)-\mu_{\alpha}\mathcal{N}^{}_{\alpha}\right)/k_{B}T_{\alpha}\right]}
{\text{tr}_{\text{}}\exp\left[-\left(\mathcal{H}^{}_{\alpha}(t_0)-\mu_{\alpha}\mathcal{N}^{}_{\alpha}\right)/k_{B}T_{\alpha}\right]},
\end{align}
\end{subequations} with $\mathcal{H}_{\alpha}(t)=\sum_{k\in\alpha}\varepsilon_{\alpha k}(t)c_{\alpha k}^{\dagger}c^{}_{\alpha k}$, $\mathcal{N}^{}_{\alpha}=\sum_{k\in\alpha}c_{\alpha k}^{\dagger}c^{}_{\alpha k}$, and $k_{B}$, the Boltzmann constant. The initial state of the principal system, $\hat{\rho}^{}_{}(t_0)$, is arbitrary.

The exact equation of motion describing the time evolution of the reduced density operator is readily given by,\cite{Tu08235311,Jin10083013}
\begin{subequations}
\label{rhocrnt}
\begin{align}
&\frac{d}{dt}\hat{\rho}(t)=-i[\mathcal{H}_{\mathrm{S}}(t),\hat{\rho}(t)]+\sum_{\alpha}[%
\mathcal{L}^{+}_{\alpha}(t)+\mathcal{L}^{-}_{\alpha}(t)]\hat{\rho}(t),
\label{rho}
\end{align}
where the superoperators $\mathcal{L}^{\pm}_{\alpha}(t)$ are
expressed explicitly by
\end{subequations}
\begin{align}
\mathcal{L}^{+}_{\alpha}(t)\hat{\rho}(t)=-\sum_{ij}\Big\{&{\boldsymbol
\lambda}_{\alpha
ij}(t)\big[a^{\dag}_{i}a_{j}\hat{\rho}(t)+a^{\dag}_{i}\hat{\rho}(t)a_{j}\big]
\nonumber\\&
+{\ \boldsymbol \kappa}_{\alpha ij}(t)a^{\dag}_{i}a_{j}\hat{\rho}(t)+%
\mathrm{h.c.}\Big\},  \notag \\
\mathcal{L}^{-}_{\alpha}(t)\hat{\rho}(t)=\sum_{ij}\Big\{&{\boldsymbol
\lambda}_{\alpha
ij}(t)\big[a_{j}\hat{\rho}(t)a^{\dag}_{i}+\hat{\rho}(t)a_{j}a^{\dag}_{i}\big]
\nonumber\\&
+{ \boldsymbol \kappa}_{\alpha ij}(t)a_{j}\hat{\rho}(t)a^{\dag}_{i}+%
\mathrm{h.c.}\Big\} ,  \label{superoperators}
\end{align}
The time-dependent dissipation and fluctuation coefficients in Eqs.~(\ref%
{superoperators}), $\boldsymbol \kappa_{\alpha}(t)$ and
${\boldsymbol \lambda}_{\alpha}(t)$, are explicitly determined by
the elementary functions, $\boldsymbol u(\tau,t_{0})$, $\boldsymbol{
\bar{u}}(\tau,t)$ and $\boldsymbol v(\tau,t)$, via the relations
\begin{subequations}
\label{kl}
\begin{align}
\boldsymbol \kappa_{\alpha}(t)=&\int_{t_0}^{t}d\tau {\boldsymbol g}_{\alpha}(t,\tau){\boldsymbol u}%
(\tau,t_0){\boldsymbol u}^{-1}(t,t_0) , \\
{\boldsymbol \lambda}_{\alpha}(t)=&\int_{t_0}^{t}\!\!\!d\tau\left\{{\boldsymbol g}%
_{\alpha}(t,\tau){\boldsymbol v}(\tau,t)-\widetilde{\boldsymbol g}_{\alpha}(t,\tau)\bar{\boldsymbol u}%
(\tau,t)\right\}\nonumber\\&-\boldsymbol \kappa_{\alpha}(t){\bm v}(t,t) .
\end{align}
These elementary functions obey the following
dissipation-fluctuation integrodifferential equations of motion
\end{subequations}
\begin{subequations}
\label{t-uvn}
\begin{equation}
\frac{\partial}{\partial\tau }\boldsymbol{u}\left( \tau,s \right)
+i\boldsymbol{E}\left( \tau \right) \boldsymbol{u}\left( \tau,s\right) +\int_{s}^{\tau }d\tau
^{\prime }\boldsymbol{g}\left( \tau ,\tau ^{\prime }\right) \boldsymbol{u}%
\left( \tau ^{\prime },s\right) =0, \label{prog_u}
\end{equation} for $t_0\leq s \leq \tau \leq t$,
\begin{align}
\boldsymbol{v}\left( \tau,t \right)
=\int_{t_{0}}^{\tau}ds\int_{t_{0}}^{t}ds^{\prime
}\boldsymbol{u}\left( \tau,s \right)\widetilde{\boldsymbol{g}}\left( s ,s ^{\prime }\right)
\boldsymbol{\bar{u}}\left( s^{\prime },t\right)\label{prog_v},
\end{align}
\end{subequations} and $\boldsymbol{\bar{u}}(\tau,t)=[\boldsymbol{u}\left( t,\tau \right)]^{\dagger}$
with $\boldsymbol u(s,s)=\mathbf{1}_{D}$. They are directly related to the nonequililbrium Green functions.\cite{Jin10083013} Here $[\boldsymbol{E}\left( \tau \right)]_{ij}=\varepsilon^{}_{ij}(\tau)$ and
\begin{subequations}
\label{t-gtg}
\begin{align}
{\boldsymbol g}(t_{1},t_{2})=\sum_{\alpha}{\boldsymbol
g}_{\alpha}(t_{1},t_{2}),~ \widetilde{\boldsymbol{
g}}(t_{1},t_{2})=\sum_{\alpha}\widetilde{\boldsymbol{g}}_{\alpha}(t_{1},t_{2})
\end{align}
with
\begin{eqnarray}
 {\boldsymbol g}_{\alpha}(t_{1},t_{2})  &=& \int \frac{d\omega
}{2\pi } \boldsymbol{\Gamma }_{\alpha }\left( \omega
,t_{1},t_{2}\right) e^{-i\omega \left( t_{1}-t_{2}\right) },\text{ }\label{dissip-kernel} \\
\widetilde{\boldsymbol{g}}_{\alpha}(t_{1},t_{2}) &=& \int
\frac{d\omega }{2\pi }f_{\alpha}(\omega) \boldsymbol{\Gamma
}_{\alpha }\left( \omega ,t_{1},t_{2}\right)e^{-i\omega \left(
t_{1}-t_{2}\right) }\label{fluc-kernel},
\end{eqnarray}
are the reservoir correlation functions and $f_{\alpha}(\omega)$ is the fermi distribution function for the initial equilibrium of the reservoir $\alpha$. The time-dependent spectral density (level-broadening) function is given by
\begin{align}
&\left[ \boldsymbol{\Gamma }_{\alpha }\left( \omega ,t_{1}
,t_{2}\right) \right] _{ij}=
\nonumber\\&
2\pi \sum_{\bm{k}\in \alpha }V_{i\alpha
\bm{k}}\left( t_{1} \right) e^{-i\int_{t_{2}
}^{t_{1}}ds\varepsilon _{\alpha \bm{k}}\left( s\right)} V_{\alpha
\bm{k}j}\left( t_{2}\right) \delta \left( \omega -\varepsilon _{\alpha
\bm{k}}(t_{0})\right).\label{time-dep-lvl-brd}
\end{align}
\end{subequations}




\subsection{General analytical solution}
\label{sec_sol}
The above master equation was derived using the Feynman-Vernon's influence functional approach in the fermion-coherent-state representation. A matrix element of the reduced density operator in the fermion-coherent-state representation is expressed as\cite{Tu08235311,Jin10083013}
\begin{subequations}
\label{rho_J_xi}
\begin{align}
\label{def_rho_xi}
\langle\boldsymbol{\xi}_{f}\vert\hat{\rho}(t)\vert\boldsymbol{\xi}'_{f}\rangle\!\!=\!\!\!\!\int\!\!d\mu(\boldsymbol{\xi}_{0})d\mu(\boldsymbol{\xi}'_{0})
\!J(\bar{\boldsymbol{\xi}}_{f},\boldsymbol{\xi}'_{f},t\vert\boldsymbol{\xi}_{0},\bar{\boldsymbol{\xi}}'_{0})
\!\langle\boldsymbol{\xi}_{0}\vert\hat{\rho}(t_0)\vert\boldsymbol{\xi}'_{0}\rangle,
\end{align} where the fermion coherent states are given by $a^{}_{i}\vert\boldsymbol{\xi}\rangle=\xi^{}_{i}\vert\boldsymbol{\xi}\rangle$ and $\langle\boldsymbol{\xi}\vert a^{\dagger}_{i}=\langle\boldsymbol{\xi}\vert\xi^{*}_{i}$ with $\boldsymbol{\xi}=\left(\xi^{}_{1},\xi^{}_{2},\cdots,\xi^{}_{D}\right)^{T}$ being a column vector of Grassman numbers $\{\xi^{}_{i}\}_{i=1}^{D}$ and $\bar{\boldsymbol{\xi}}=\left(\xi^{*}_{1},\xi^{*}_{2},\cdots,\xi^{*}_{D}\right)$ being a row vector. The propagating function in the fermion-coherent-state representation is given by
\begin{align}
\label{def_J_xi_withF}
&J(\bar{\boldsymbol{\xi}}_{f},\boldsymbol{\xi}'_{f},t\vert\boldsymbol{\xi}_{0},\bar{\boldsymbol{\xi}}'_{0})=\nonumber\\
&
\int\mathcal{D}\left[\bar{\boldsymbol{\xi}},\boldsymbol{\xi};\bar{\boldsymbol{\xi}}',\boldsymbol{\xi}'\right]
e^{i\left(\mathcal{S}_{c}[\bar{\boldsymbol{\xi}},\boldsymbol{\xi}]-\mathcal{S}_{c}[\bar{\boldsymbol{\xi}}',\boldsymbol{\xi}']\right)}
\mathcal{F}[\bar{\boldsymbol{\xi}},\boldsymbol{\xi};\bar{\boldsymbol{\xi}}',\boldsymbol{\xi}'],
\end{align}
in which $\mathcal{S}_{c}[\bar{\boldsymbol{\xi}},\boldsymbol{\xi}]$ is the action of the principal system and $\mathcal{F}[\bar{\boldsymbol{\xi}},\boldsymbol{\xi};\bar{\boldsymbol{\xi}}',\boldsymbol{\xi}']$ is the influence functional obtained by integrating out the reservoirs' degrees of freedom (see Refs.~[\onlinecite{Tu08235311,Jin10083013}] for the explicit derivations). Specifying the action of the principal system with Eq.~(\ref{gel_Hs}), the Grassmann number integral in Eq.~(\ref{def_J_xi_withF}) has been exactly carried out, yielding,
\begin{align}
&J(\bar{\boldsymbol{\xi}}_{f},\boldsymbol{\xi}'_{f},t\vert\boldsymbol{\xi}_{0},\bar{\boldsymbol{\xi}}'_{0})=\det\left[\mathbf{1}_{D}-\boldsymbol{v}\left(t\right)\right]
\times
\nonumber\\
&\exp\!\left(\bar{\boldsymbol{\xi}}_{f}\boldsymbol{J}_{f0}(t)\boldsymbol{\xi}_{0}\!\!+\!\!\bar{\boldsymbol{\xi}}'_{0}\boldsymbol{J}_{0f}(t)\boldsymbol{\xi}'_{f}
\!\!+\!\!\bar{\boldsymbol{\xi}}_{f}\boldsymbol{J}_{ff}(t)\boldsymbol{\xi}'_{f}\!\!+\!\!\bar{\boldsymbol{\xi}}'_{0}\boldsymbol{J}_{00}(t)\boldsymbol{\xi}_{0} \right),
\end{align} in which we have introduced $D\times D$ matrices,
\begin{align}
\label{specifc-J_bn}
&\boldsymbol{J}_{f0}\left( t\right)
=[\boldsymbol{J}_{0f}\left( t\right)]^{\dagger}=(\mathbf{1}_{D}-\boldsymbol{v}(t))^{-1}\boldsymbol{u}(t),
\nonumber\\&
\boldsymbol{J}_{00}\left( t\right) =\boldsymbol{u}^{\dag }(t)(\mathbf{1}_{D}-\boldsymbol{v}(t))^{-1}
\boldsymbol{u}(t)-\mathbf{1}_{D},
\nonumber\\&
\boldsymbol{J}_{ff}\left( t\right) =(\mathbf{1}_{D}-\boldsymbol{v}(t))^{-1}-\mathbf{1}_{D},
\end{align} with $\boldsymbol{u}(t)=\boldsymbol{u}(t,t_0)$ and $\boldsymbol{v}(t)=\boldsymbol{v}(t,t)$.
\end{subequations} Since fermion coherent states are used only as a mathematical tool, it is not straightforward to gain understanding of the physical properties of the resulting quantum states $\hat{\rho}(t)$ directly from Eq.~(\ref{rho_J_xi}). To meaningfully express the reduced density operator, we have to work in the Fock-state representation.

The Fock basis is defined by
\begin{align}
\label{gel_dstmtx_notat}
\left.\vert n_{1},n_{2},\cdots,n_{D}\rangle\right.=(a^{\dagger}_{1})^{n_{1}}(a^{\dagger}_{2})^{n_{2}}\cdots(a^{\dagger}_{D})^{n_{D}}\vert0\rangle,
\end{align}
where $\vert0\rangle$ is the configuration where the principal system is completely unoccupied. Here $\{n_{i}\}_{i=1}^{D}\in\left\{ 0,1\right\}$ enumerates the number of particles occupying the $i$th orbital in the principal system, for $i\in\text{S}$. However, when $D$ is arbitrary, it is not convenient to denote these states using a set of $D$ integers as done in Eq.~(\ref{gel_dstmtx_notat}). An $n$-particle configuration is specified by a certain set of $n$ occupied orbitals, $\boldsymbol{i}^{(n)}=\{i_{1},\cdots,i_{n}\}$, in which $i_{1},\cdots,i_{n}\in\text{S}$. Ambiguity of defining its corresponding $n$-particle state arises from the anti-commutating property of the fermion field operators, namely, $a^{\dagger}_{i_{n}}\cdots a^{\dagger}_{i_{3}}a^{\dagger}_{i_{2}}a^{\dagger}_{i_{1}}\vert0\rangle=-a^{\dagger}_{i_{n}}\cdots a^{\dagger}_{i_{3}}a^{\dagger}_{i_{1}}a^{\dagger}_{i_{2}}\vert0\rangle$. To avoid this ambiguity, we enforce an ordering regulation to all many-particle states using the notation,
\begin{align}
\label{ordered_fock_state}
\vert P\boldsymbol{i}^{(n)} \rangle
=a^{\dagger}_{P\boldsymbol{i}^{(n)}_{n}}\cdots a^{\dagger}_{P\boldsymbol{i}^{(n)}_{1}}\vert 0 \rangle,
\end{align} where $P$ stands for an orderer that $P\boldsymbol{i}^{(n)}=\{P\boldsymbol{i}^{(n)}_{n},\cdots,P\boldsymbol{i}^{(n)}_{2}, P\boldsymbol{i}^{(n)}_{1}\}$ is an ordered set, in which all the entries $P\boldsymbol{i}^{(n)}_{n},\cdots,P\boldsymbol{i}^{(n)}_{2}, P\boldsymbol{i}^{(n)}_{1}\in\boldsymbol{i}^{(n)}$, are with the ordering $P\boldsymbol{i}^{(n)}_{n}>\cdots>P\boldsymbol{i}^{(n)}_{2}>P\boldsymbol{i}^{(n)}_{1}$. With Eq.~(\ref{ordered_fock_state}), we then unambiguously write down the elements of the reduced density matrix as
\begin{align}
\label{RDM_elem_expm}
\left.\langle P\boldsymbol{i}^{(n)}\vert\right.\hat{\rho}_{}\left(t\right)\left.\vert P\boldsymbol{j}^{(n)}\rangle\right.,
\end{align} to represent the coherence between the configuration $\boldsymbol{i}^{(n)}$ and $\boldsymbol{j}^{(n)}$ if $\boldsymbol{i}^{(n)}\ne\boldsymbol{j}^{(n)}$ or the probability for the system to be in the configuration $\boldsymbol{i}^{(n)}$ if $\boldsymbol{i}^{(n)}=\boldsymbol{j}^{(n)}$.

Using the transformation between the Fock states and the fermion coherent states,\cite{Faddeev80book}
\begin{equation}
\langle P\boldsymbol{i}^{(n)}|\boldsymbol{\xi }\rangle =\xi
_{P\boldsymbol{i}^{(n)}_{1}}\xi_{P\boldsymbol{i}^{(n)}_{2}}\cdot \cdot \cdot \xi
_{P\boldsymbol{i}^{(n)}_{n}},
\end{equation} Eq.~(\ref{rho_J_xi}) can be cast into the Fock-state representation for arbitrary $D$ and arbitrary initial state of the principal system.
This reads,
\begin{subequations}
\label{densitymatrix-gel-sol}
\begin{align}
\label{dsty_t_0_bytotprog}
&\langle P\boldsymbol{l}^{(n)}\vert\hat{\rho}_{}
(t)\vert P\boldsymbol{r}^{(n)}\rangle=\det\left[\mathbf{1}_{D}-\boldsymbol{v}
\left( t\right) \right]\nonumber\\&\times\sum_{m=0}^{D}\sum_{P\boldsymbol{a}^{(m)}}^{}\sum_{P\boldsymbol{b}^{(m)}}^{}\Big[
\mathcal{J}_{\boldsymbol{a}^{(m)},\boldsymbol{b}^{(m)}}^{\boldsymbol{l}^{(n)},\boldsymbol{r}^{(n)}}(t)\langle P\boldsymbol{a}^{(m)}\vert\hat{\rho}_{} (t_0)\vert P\boldsymbol{b}^{(m)}\rangle \Big],
\end{align}%
where $\det\left[\mathbf{1}_{D}-\boldsymbol{v}
\left( t\right) \right]\mathcal{J}_{\boldsymbol{a}^{(m)},\boldsymbol{b}^{(m)}}^{\boldsymbol{l}^{(n)},\boldsymbol{r}^{(n)}}(t)$ connects the initial value of the reduced density matrix evaluated between the configurations $\boldsymbol{a}^{(m)}$ and $\boldsymbol{b}^{(m)}$ to the later value of the reduced density matrix evaluated between the configurations $\boldsymbol{l}^{(n)}$ and $\boldsymbol{r}^{(n)}$. Explicitly, they read
\begin{align}
\label{total_prog}
&\mathcal{J}_{\boldsymbol{e}^{(0)},\boldsymbol{e}^{(0)}}^{\boldsymbol{e}^{(0)},\boldsymbol{e}^{(0)}}(t)=1
,\nonumber\\&
\mathcal{J}_{\boldsymbol{e}^{(0)},\boldsymbol{e}^{(0)}}^{\boldsymbol{l}^{(n\ne0)},\boldsymbol{r}^{(n\ne0)}}(t)=
\det\left(\left[\boldsymbol{J}_{ff}\left(t\right)\right]_{\boldsymbol{l}^{(n)}}^{\boldsymbol{r}^{(n)}}\right),
\nonumber\\&
\mathcal{J}_{\boldsymbol{a}^{(m\ne0)}\!\!,\boldsymbol{b}^{(m\ne0)}}^{\boldsymbol{e}^{(0)},\boldsymbol{e}^{(0)}}\!(t)\!=\!
(-1)^{m}\det\left(\left[\boldsymbol{J}_{00}\left(t\right)\right]_{\boldsymbol{b}^{(m)}}^{\boldsymbol{a}^{(m)}}\right),
\nonumber\\
&\mathcal{J}_{\boldsymbol{a}^{(m\ne0)}\!,\boldsymbol{b}^{(m\ne0)}}^{\boldsymbol{l}^{(n\ne0)},\boldsymbol{r}^{(n\ne0)}}\!(t)\!=\!
(-1)^{m}\!\det\!\!\left(\!\!
\begin{array}{ll}
\left[\boldsymbol{J}_{00}\left(t\right)\right]_{\boldsymbol{b}^{(m)}}^{\boldsymbol{a}^{(m)}}
& \left[\boldsymbol{J}_{0f}\left(t\right)\right]_{\boldsymbol{b}^{(m)}}^{\boldsymbol{r}^{(n)}}\\
\left[\boldsymbol{J}_{f0}\left(t\right)\right]_{\boldsymbol{l}^{(n)}}^{\boldsymbol{a}^{(m)}}
& \left[\boldsymbol{J}_{ff}\left(t\right)\right]_{\boldsymbol{l}^{(n)}}^{\boldsymbol{r}^{(n)}}
\end{array}\!\!
\right),
\end{align} where $\boldsymbol{e}^{(0)}$ represents the vacuum state.
 The notations $\left[\boldsymbol{J}_{00}\left(t\right)\right]_{\boldsymbol{b}^{(m)}}^{\boldsymbol{a}^{(m)}}$, $\left[\boldsymbol{J}_{0f}\left(t\right)\right]_{\boldsymbol{b}^{(m)}}^{\boldsymbol{r}^{(n)}}$, $\left[\boldsymbol{J}_{f0}\left(t\right)\right]_{\boldsymbol{l}^{(n)}}^{\boldsymbol{a}^{(m)}}$ and $\left[\boldsymbol{J}_{ff}\left(t\right)\right]_{\boldsymbol{l}^{(n)}}^{\boldsymbol{r}^{(n)}}$ in Eq.~(\ref{total_prog}) are $m\times m$, $m\times n$, $n\times m$ and $n\times n$ matrices respectively, whose elements are given by
\begin{align}
\label{orbt-slect}
&\left(\left[\boldsymbol{J}_{00}\left(t\right)\right]_{\boldsymbol{b}^{(m)}}^{\boldsymbol{a}^{(m)}}\right)_{ij}=\left[
\boldsymbol{J}_{00}\left( t\right) \right]
_{P\boldsymbol{b}^{(m)}_{i}P\boldsymbol{a}^{(m)}_{j}},\nonumber\\
&\left(\left[\boldsymbol{J}_{0f}\left(t\right)\right]_{\boldsymbol{b}^{(m)}}^{\boldsymbol{r}^{(n)}}\right)_{ij}=
\left[
\boldsymbol{J}_{0f}\left( t\right) \right]_{P\boldsymbol{b}^{(m)}_{i}P\boldsymbol{r}^{(n)}_{j}},\nonumber\\
&\left(\left[\boldsymbol{J}_{f0}\left(t\right)\right]_{\boldsymbol{l}^{(n)}}^{\boldsymbol{a}^{(m)}}\right)_{ij}
=\left[\boldsymbol{J}_{f0}\left(t\right)\right]_{P\boldsymbol{l}^{(n)}_{i}P\boldsymbol{a}^{(m)}_{j}},
\nonumber\\
&\left(\left[\boldsymbol{J}_{ff}\left(t\right)\right]_{\boldsymbol{l}^{(n)}}^{\boldsymbol{r}^{(n)}}\right)_{ij}
=\left[\boldsymbol{J}_{ff}\left(t\right)\right]_{P\boldsymbol{l}^{(n)}_{i}P\boldsymbol{r}^{(n)}_{j}},
\end{align} for $i,j$ being in their respective ranges.
\end{subequations}

Equation (\ref{densitymatrix-gel-sol}) is the general analytical solution to the exact fermion master equation.
Note that due to the conservation of the total particle number, the
dynamics for certain elements of the reduced density operator, $\langle
P\boldsymbol{l}^{(n)}|\hat{\rho}(t)|P\boldsymbol{b}^{(m)}\rangle$ with
$n\ne m$ is decoupled from those with $n=m$.  Since there
is no pair creation or pair annihilation present in the Hamiltonian of
Eq.~(\ref{gel_H}), we are not interested in these
matrix elements resulted from superposition between configurations of different particle numbers.  For physically
meaningful preparations of the initial states, the elements $\langle
P\boldsymbol{l}^{(n)}|\hat{\rho}(t_0)|P\boldsymbol{b}^{(m)}\rangle$
with $n\ne m$ are zero for all configurations
$\boldsymbol{l}^{(n)}$ and $\boldsymbol{b}^{(m)}$ and all values
of $n$ and $m$.  The subsequent time evolution therefore guarantees
$\langle P\boldsymbol{l}^{(n)}|\hat{\rho}
(t\ge t_{0})|P\boldsymbol{b}^{(m)}\rangle=0$ for any $\boldsymbol{l}^{(n)}$ and $\boldsymbol{b}^{(m)}$ with $n\ne m$.

\section{Example: Transient dynamics of artificial molecules}
\label{sec_DQDmodel}

In this section, we demonstrate the convenience of using the above solution with an example, namely, the dynamics of an artificial molecule formed in electronic nanostructures or its atomic analog.  It consists of two atomic orbitals (localized orbital in a QD or a potential well) tunnel coupled to each other. This corresponds to set $D=2$ in Eq.~(\ref{gel_Hs}). We denote the possible configuration of occupying the atomic orbitals by $\vert0\rangle$ (both atomic orbitals are not occupied), $\vert1\rangle$ (orbital 1 is occupied and orbital 2 is empty), $\vert2\rangle$ (orbital 2 is occupied and orbital 1 is empty), and $\vert3\rangle$ (both atomic orbitals are occupied).

The expression for the reduced density operator in the Fock basis, Eq.~(\ref{densitymatrix-gel-sol}), then explicitly reads,
\begin{subequations}
\label{dsytmx-D2}
\begin{equation}
\langle i\vert\hat{\rho}(t)\vert j\rangle=\rho_{ij}\left(t\right)=\rho_{ij}^{{\rm init.}}\left(t\right)+\rho_{ij}^{{\rm em.}}(t),
\end{equation} for $i,j\in\{0,1,2,3\}$. The part that does not depend on the initial state of the principal system is
\begin{align}
\rho_{33}^{{\rm em.}}(t) & =  {\rm det}\left[\boldsymbol{v}\left(t\right)\right],\nonumber\\
\rho_{11}^{{\rm em.}}(t) & =  v_{11}\left(t\right)-{\rm det}\left[\boldsymbol{v}\left(t\right)\right],\nonumber\\
\rho_{22}^{{\rm em.}}(t) & =  v_{22}\left(t\right)-{\rm det}\left[\boldsymbol{v}\left(t\right)\right],\nonumber\\
\rho_{00}^{{\rm em.}}(t) & =  1-\left(v_{11}\left(t\right)+v_{22}\left(t\right)\right)+{\rm det}\left[\boldsymbol{v}\left(t\right)\right],\nonumber\\
\rho_{21}^{{\rm em.}}(t) & =  v_{21}\left(t\right).
\end{align} The part that contains the initial state is given by
\begin{widetext}
\begin{align}
\rho_{33}^{{\rm init.}}\left(t\right) =& v_{11}\left(t\right)\varrho_{22}^{0}\left(t\right)+v_{22}\left(t\right)\varrho_{11}^{0}\left(t\right)
-v_{12}\left(t\right)\varrho_{21}^{0}\left(t\right)-v_{21}\left(t\right)\varrho_{12}^{0}\left(t\right)
\nonumber \\
&+\rho_{33}\left(t_0\right){\rm det}\left[\boldsymbol{u}^{\dagger}\left(t\right)\boldsymbol{u}\left(t\right)\right],
\nonumber\\
\rho_{11}^{{\rm init.}}\left(t\right)  =&  \left(1-v_{22}\left(t\right)\right)\varrho_{11}^{0}\left(t\right)-v_{11}\left(t\right)\varrho_{22}^{0}\left(t\right)
+v_{12}\left(t\right)\varrho_{21}^{0}\left(t\right)+v_{21}\left(t\right)\varrho_{12}^{0}\left(t\right)
\nonumber \\
&-\rho_{33}\left(t_0\right){\rm det}\left[\boldsymbol{u}^{\dagger}\left(t\right)\boldsymbol{u}\left(t\right)\right],
\nonumber \\
\rho_{22}^{{\rm init.}}\left(t\right)  =&  \left(1-v_{11}\left(t\right)\right)\varrho_{22}^{0}\left(t\right)-v_{22}\left(t\right)\varrho_{11}^{0}\left(t\right)
+v_{12}\left(t\right)\varrho_{21}^{0}\left(t\right)+v_{21}\left(t\right)\varrho_{12}^{0}\left(t\right)
\nonumber \\
&-\rho_{33}\left(t_0\right){\rm det}\left[\boldsymbol{u}^{\dagger}\left(t\right)\boldsymbol{u}\left(t\right)\right],
\nonumber \\
\rho_{00}^{{\rm init.}}\left(t\right)=&  -\left[\left(1-v_{22}\left(t\right)\right)\varrho_{11}^{0}\left(t\right)+\left(1-v_{11}\left(t\right)\right)\varrho_{22}^{0}\left(t\right)
+v_{12}\left(t\right)\varrho_{21}^{0}\left(t\right)+v_{21}\left(t\right)\varrho_{12}^{0}\left(t\right)\right]
\nonumber\\
&+\rho_{33}\left(t_0\right){\rm det}\left[\boldsymbol{u}^{\dagger}\left(t\right)\boldsymbol{u}\left(t\right)\right],
 \nonumber \\
\rho_{21}^{{\rm init.}}\left(t\right)  =& \varrho_{21}^{0}\left(t\right),
\end{align}
\end{widetext} where the dependencies on the initial state of the molecule appear explicitly with $\rho_{33}\left(t_0\right)$ and
\begin{equation}
\varrho_{ij}^{0}\left(t\right)=\sum_{l,l'\in\text{S}}
[\boldsymbol{u}\left(t\right)]_{il}\text{tr}[a_{l}\hat{\rho}(t_{0})a^{\dagger}_{l'}]\left[\boldsymbol{u}\left(t\right)\right]^{\dagger}_{l'j}.
\end{equation}
\end{subequations} We have denoted $v_{ij}(t)=[\boldsymbol{v}\left(t\right)]_{ij}$ for $i,j\in\{1,2\}$. In case that the initial state of the principal system is an empty state, then Eq.~(\ref{dsytmx-D2}) becomes $\rho_{ij}^{}\left(t\right)=\rho_{ij}^{{\rm em.}}\left(t\right)$, reproducing the solution in Ref.~[\onlinecite{Tu11115318}].


In what follows, we concentrate on the role of the initial states on the dynamics of the molecule. We consider the situation where the total Hamiltonian is independent of time. Then the spectral density in Eq.~(\ref{t-gtg}) becomes independent of time, namely,
\begin{align}
\label{lvbrd-indt}
\left[ \boldsymbol{\Gamma }_{\alpha }\left( \omega ,t_{1}
,t_{2}\right) \right] _{ij}=\Gamma^{\alpha}_{ij}(\omega)=2\pi\sum_{k\in\alpha}\delta(\omega-\varepsilon_{\alpha k})V_{i\alpha k}V_{\alpha kj}.
\end{align}
We use the widely applied wide-band limit, leading Eq.~(\ref{lvbrd-indt}) to $\Gamma^{\alpha}_{ij}(\omega)=\Gamma^{\alpha}_{ij}$. In this case, $\lim_{t\rightarrow\infty}\boldsymbol{u}\left(t\right)=0$, and we immediately see from Eq.~(\ref{dsytmx-D2}) that the part which carries the information of the initial states of the molecule, vanishes in the steady limit
\begin{align}
\lim_{t\rightarrow\infty}\rho_{ij}^{{\rm init.}}\left(t\right)=0.
\end{align} The initial state of the molecule only has a transient effect. We also assume there is only one reservoir so that the label $\alpha$ in Eq.~(\ref{lvbrd-indt}) can be removed.

The molcular bonding and anti-bonding states (BS and AS) are obtained with equal on-site energies $\varepsilon_{11}=\varepsilon_{22}=\varepsilon_{0}$. In this basis, the Hamiltonian of the molecule becomes
\begin{subequations}
\begin{align}
\mathcal{H}_{S}=\Omega(d^{\dagger}_{\text{AS}}d^{}_{\text{AS}}-d^{\dagger}_{\text{BS}}d^{}_{\text{BS}}),
\end{align} where the molecular orbitals in terms of atomic orbitals are
\begin{align}
&d^{\dagger}_{\text{BS}}=\frac{1}{\sqrt{2}}(a^{\dagger}_{1}+a^{\dagger}_{2}),
\nonumber\\&
d^{\dagger}_{\text{AS}}=\frac{1}{\sqrt{2}}(a^{\dagger}_{1}-a^{\dagger}_{2}),
\end{align} and $\Omega=\vert\varepsilon_{21}\vert$ is the bonding strength between the two artificial atoms.
\end{subequations} The probabilities $\langle B\vert\hat{\rho}(t)\vert B\rangle=\rho^{}_{BB}(t)$ and $\langle A\vert\hat{\rho}(t)\vert A\rangle=\rho^{}_{AA}(t)$, where $\vert B\rangle=d^{\dagger}_{\text{BS}}\vert 0\rangle$ and $\vert A\rangle=d^{\dagger}_{\text{AS}}\vert 0\rangle$, tell to which extent the principal system is in a well-defined molecular state.

There are two categories of transition path ways that lead the changes of the molecular quantum states. One is the internal transition induced by coherent coupling between the two atomic orbitals. The other is via exchanging particles with the external reservoir. By preparing the molecule at different initial states, we are able to unfold the actions of these different transition pathways.

\begin{figure}[h]
\includegraphics[width=8.2cm,
height=6.0cm]{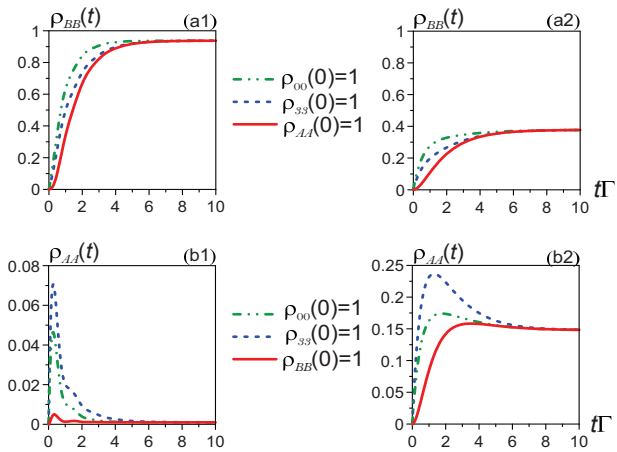} \caption{(color online) The left panel (plots (a1) and (b1)) is for $\Omega=5\Gamma$ and the right panel (plots (a2) and (b2)) is for $\Omega=\Gamma/5$. Plots (a1) and (a2) show how $\rho_{BB}^{}(t)$ evolves in time starting various initial states as labeled. Plots (b1) and (b2) show the time evolutions of $\rho_{AA}^{}(t)$. The parameters used here and also in other figures are $\mu=\varepsilon_{0}$ and $k_{B}T=0.1\Gamma$ with $\Gamma_{22}/\Gamma_{11}=0.7$ and $\Gamma_{12}=0.2\sqrt{\Gamma_{11}\Gamma_{22}}$.} \label{fig2}
\end{figure}

In Fig.~\ref{fig2} we respectively monitor the time evolution of probabilities for the one-particle BS and AS, starting from various initial states of the molecule.
The situation where BS and AS are well separated in energy, namely, $\Omega>2\Gamma$, where $\Gamma=(\Gamma_{11}+\Gamma_{22})$/2 is the average level broadening, is shown on the left panel. The situation that their separation is within the broadening $\Omega<\Gamma$ is displayed on the right panel. Since BS and AS are the eigenstates of the bare molecule, the transition $\vert A\rangle\rightarrow\vert B\rangle$ occur only via exchanging particles with the reservoir, through $\vert A\rangle\rightarrow\vert 0\rangle\rightarrow\vert B\rangle$ and $\vert A\rangle\rightarrow\vert 3\rangle\rightarrow\vert B\rangle$. This is true also for the transition $\vert B\rangle\rightarrow\vert A\rangle$. Therefore the resulting probability $\rho^{}_{BB}(t)$ ($\rho^{}_{AA}(t)$) obtained by starting the system from the state $\vert A\rangle$ ($\vert B\rangle$) in the transient process is not as high as those obtained by starting the system from $\vert 0\rangle$ or $\vert 3\rangle$. This is shown in Fig.~\ref{fig2}, where the value of the red solid curves (for $\rho_{BB}(t)$ starting from AS in Fig.~\ref{fig2}(a1),(a2) and for $\rho_{AA}(t)$ starting from BS in Fig.~\ref{fig2}(b1),(b2)) are transiently smaller than the other curves (for those starting from empty state or two-particle state). When AS and BS are well separated in energy with the chemical potential of the reservoir being placed below AS but above BS, then the molecule tends to reside in the BS, leaving negligible value of $\rho_{AA}(t)$. Such preference is shown in Fig.~\ref{fig2}(a1) (with larger $\rho_{BB}(t)$) and in Fig.~\ref{fig2}(a2) (with smaller $\rho_{BB}(t)$). The probability for AS can be enhanced by reducing $\Omega$ to be smaller than the broadenings (see $\rho^{}_{AA}(t)$ in Fig.~\ref{fig2}(b2) being larger than those in Fig.~\ref{fig2}(b1)).

\begin{figure}[h]
\includegraphics[width=8.2cm,
height=6.0cm]{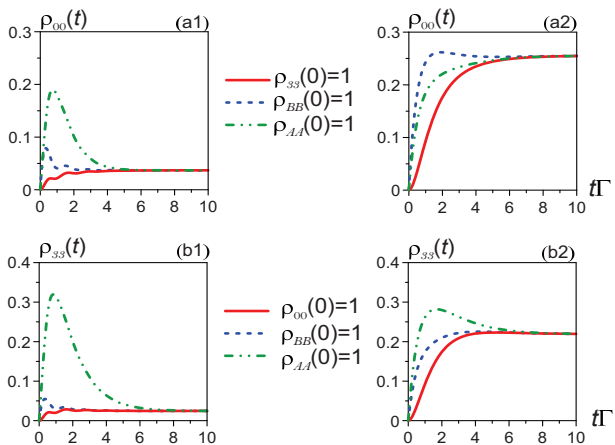} \caption{(color online) Plots (a1) and (a2) show the time evolutions of $\rho_{00}^{}(t)$ initiated from various states as labeled. Plots (b1) and (b2) display the time evolutions of $\rho_{33}^{}(t)$. In plots (a1) and (b1) we let $\Omega=5\Gamma$. In In plots (a2) and (b2) we let $\Omega=\Gamma/5$} \label{fig3}
\end{figure}

The probabilities for empty state and two-particle state, starting from other initial states, are shown in Fig.~\ref{fig3}. The transitions $\vert 0\rangle\rightarrow\vert 3\rangle$ and $\vert 3\rangle\rightarrow\vert 0\rangle$ involve intermediate one-particle states, $\vert 0\rangle\rightarrow\vert B(A)\rangle\rightarrow\vert 3\rangle$ and $\vert 3\rangle\rightarrow\vert B(A)\rangle\rightarrow\vert 0\rangle$. Therefore it is less efficient to reach $\vert 0\rangle$ ($\vert 3\rangle$) from $\vert 3\rangle$ ($\vert 0\rangle$) than to reach it by starting the system from the one-particle states $\vert B\rangle$ and $\vert A\rangle$. This is shown in Fig.~\ref{fig3}, where the values of red solid curves are transiently smaller than that of the others. Since the one-particle BS is less preferable in case of $\Omega<\Gamma$ than the case with $\Omega>2\Gamma$, the probability to stay out of one-particle space is relatively higher in the former case (see $\rho^{}_{00}(t)$ ($\rho^{}_{33}(t)$) in the right panel is greater than that in the left panel in Fig.~\ref{fig3} for later times).

\begin{figure}[h]
\includegraphics[width=8.2cm,
height=3.0cm]{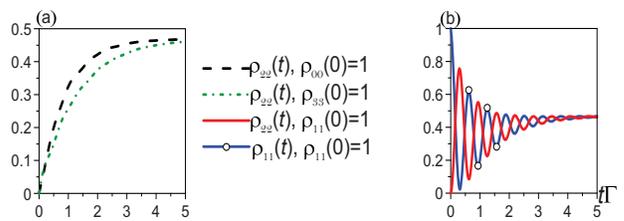} \caption{(color online) Plot (a) presents the time evolutions of $\rho_{22}(t)$ for the cases : (i) the two atomic orbitals are initially unoccupied (black long-dashed line) and (ii) both of the atomic orbitals are initially occupied (the green dash-dot line). In plot (b), the molecule starts with the one-particle state $\vert 1\rangle$. The blue solid line with marked white circles shows the subsequent time evolution of $\rho_{11}(t)$ while the red solid line is for $\rho_{22}(t)$. In this figure, we use $\Omega=5\Gamma$.} \label{fig4}
\end{figure}

The different effects due to external reservoir and due to internal coherent coupling between the atoms can be made more distinct. We monitor the dynamics of probabilities $\rho_{22}^{}(t)$ and $\rho_{11}^{}(t)$ for the localized one-particle states in Fig.~\ref{fig4}. In Fig.~\ref{fig4}(a) we show the time evolution of $\rho_{22}^{}(t)$ resulting from initializing the molecule in states $\vert0\rangle$ and $\vert3\rangle$. In contrast to these smooth evolution processes in Fig.~\ref{fig4}(a), clear oscillation of $\rho_{22}^{}(t)$ in time is observed if one starts with one particle localized in $\vert1\rangle$ (see the red solid curve in Fig.~\ref{fig4}(b)). The oscillation of $\rho_{11}^{}(t)$ starting from the same state is shown to be complementary to the oscillation of $\rho_{22}^{}(t)$ (see the blue solid line marked with white circles in Fig.~\ref{fig4}(b)). This manifests the coherent tunneling between the two atoms. We thus demonstrated the convenience of directly applying Eq.~(\ref{densitymatrix-gel-sol}) to study the transient dynamics resulted from various initial states.



\section{Summary }
\label{sec_conclu}

In conclusion, we have obtained the general analytical solution to the exact fermion master equation presented in Refs.~[\onlinecite{Tu08235311,Jin10083013}]. Such solution is given by Eq.~(\ref{densitymatrix-gel-sol}). It connects the initial state of the principal system to its state at later times. It is completely in terms of single-particle propagating and correlating Green functions given by Eqs.~(\ref{prog_u}) and (\ref{prog_v}), respectively. Direct and complete characterization of the quantum states in real time for a class of nanoelectronic systems and their cold-atom analogs is then made feasible.


Using the above solution, we have studied the transient dynamics of artificial molecules triggered by different initial states. We show that some initial states lead to more efficient occupations of the particular target states than the other initial states do. This also reflects the different pathways the molecule transverses in the presence of the dissipative contact reservoirs. By comparing the transient probabilities of reaching one particular state from different initial states, one can distinguish the internal and the external transition pathways.









%

State preparations and read outs of engineered Hamiltonian that can mimic molecules are feasible using cold atoms with magneto- and optical controls \cite{Volz06692,Mueller07200405,Bloch12267}. The readings of the state of the DQD molecule are often via separate mesoscopic electrometers,\cite{Elzerman03161308R,Gorman05090502,Petta052180,Fricke13126803} or alternatively by integrating the DQD with radio-frequency resonant circuits.\cite{Petersson102789}  The quantum state properties investigated with the example of artificial molecules here are readily to be examined in experiments.

The general formulation presented in this article is applicable to time-dependent Hamiltonian. The time-dependence, besides arising from external fields, can also come from two-body interactions approximated by suitable mean-field treatments. The general analytical solution to exact fermion master equations should have more other potential applications to various realizations of nanoscale quantum devices. An understanding of the general behaviours of the quantum states is also helpful to the development of quantum controls.

\begin{acknowledgements}
This work is partially supported by the National Science Council
(NSC) of the ROC, under Contract No. NSC102-2112-M-006-016-MY3, by the Headquarters of
University Advancement at the National Cheng Kung University, which is sponsored by the Ministry of Education, Taiwan, ROC
and  from the National Center for Theoretical Science
of NSC and the High Performance Computing Facility in the National
Cheng Kung University. We thank Prof. Aharony and Prof. Entin-Wohlman for useful discussions.
\end{acknowledgements}

\end{document}